\documentstyle[12pt,epsf]{article}
\renewcommand{\bar}[1]{\overline{#1}}

\begin{document}

\begin{flushright}
USM-TH-78
\end{flushright}

\bigskip\bigskip
\centerline{\large \bf Quark Structure of $\Lambda$ from
$\Lambda$-Polarization in Z Decays}

\vspace{22pt}

\centerline{\bf Bo-Qiang Ma$^{a}$, Ivan Schmidt$^{b}$, and
Jian-Jun Yang$^{b,c}$}

\vspace{8pt}
{\centerline {$^{a}$CCAST (World Laboratory), P.O.~Box 8730, Beijing
100080, China}}

{\centerline {and Institute of High Energy Physics, Academia Sinica,
P.~O.~Box 918(4),}

{\centerline {Beijing 100039, China\footnote{Mailing address}}

{\centerline{e-mail: mabq@hptc5.ihep.ac.cn}}

{\centerline {$^{b}$Departamento de F\'\i sica, Universidad
T\'ecnica Federico Santa Mar\'\i a,}}

{\centerline {Casilla 110-V, 
Valpara\'\i so, Chile}}

{\centerline{Email: ischmidt@fis.utfsm.cl} }

{\centerline {$^{c}$Department of Physics, Nanjing Normal
University,}}

{\centerline {Nanjing 210097, China}}

{\centerline{Email: jjyang@fis.utfsm.cl} }


\vspace{10pt}
\begin{center} {\large \bf Abstract}

\end{center}

The flavor and spin structure for the quark distributions of the
$\Lambda$-baryon is studied in a perturbative QCD (pQCD) analysis
and in  the SU(6) quark-diquark model, and then applied to
calculate the $\Lambda$-polarization of semi-inclusive $\Lambda$
production in $e^+e^-$-annihilation near the $Z$-pole. It is found
that the quark-diquark model gives very good description of the
available experimental data. The pQCD model can also give good
description of the data by taking into account the suppression of
quark helicities compared to the naive SU(6) quark model spin
distributions. Further information is required for a clean
distinction between different predictions concerning the flavor
and spin structure of the $\Lambda$.

\vfill

\centerline{PACS numbers: 14.20.Jn, 12.38.Bx, 12.39.Ki, 13.60.Hb}

\vfill
\centerline{Version to be published in Phys.~Rev.~D 61 (2000). }
\vfill
\newpage

\section{Introduction}

The flavor and spin structure of the nucleons is one of the most
active research directions of the high energy physics community.
Though there have been remarkable achievements in our knowledge of
the quark-gluon structure of the nucleons from three decades of
experimental and theoretical investigations in various deep
inelastic scattering (DIS) processes, the detailed flavor and spin
structure of nucleons remains a domain with many unknowns, and
there have been many unexpected surprises with respect to naive
theoretical considerations. The sea content of the nucleons has
received extensive investigations concerning its spin structure
\cite{SpinR}, strange content \cite{Bro88,Bro96}, flavor asymmetry
\cite{Kum97}, and isospin symmetry breaking \cite{Isospin}. Even
our knowledge of the valence quarks is still not well established,
reflected from the recent investigations concerning the flavor and
spin structure of the valence quarks for the nucleon near $x=1$.
For example, there are different predictions concerning the ratio
$d(x)/u(x)$ at $x \to 1$ from the perturbative QCD (pQCD) analysis
\cite{Far75,Bro95} and the SU(6) quark-diquark model
\cite{DQM,Ma96,Mel96}, and there are different predictions
concerning the value of $F_2^n(x)/F_2^p(x)$ at large $x$, which
has been taken to be $1/4$ as in the quark-diquark model in most
parameterizations of quark distributions. A recent analysis
\cite{Yang99} of experimental data from several processes suggests
that $F_2^n(x)/F_2^p(x) \to 3/7$ as $x \to 1$, in favor of the
pQCD prediction. The spin structure of the valence quarks is also
found to be different near $x=1$ in these models, and predictions
have been made concerning the non-dominant valence down ($d$)
quark, so that $\Delta d(x)/d(x)=-1/3$ in the quark-diquark model
\cite{Ma96,Mel96}, a result which is different from the pQCD
prediction $\Delta q(x)/q(x)=1$ for either $u$ and $d$
\cite{Bro95}. At the moment, there is still no clear data in order
to check these different predictions, although the available
measurements \cite{SMC96} for the polarized $d$ quark
distributions seem to be negative at large $x$, slightly in favor
of the quark-diquark model prediction.

It is important to perform high precision measurements of
available physical quantities and/or to measure new quantities
related to the flavor and spin structure of the nucleons, in order
to reveal more about the quark-gluon structure of the nucleons.
However, it should be more meaningful and efficient if we can find
a new domain where the same physics concerning the structure of
the nucleons can manifest itself in a way that is more easy and
clean to be detected and checked. It seems that $\Lambda$ Physics
is such a new frontier, and therefore can be used to test various
ideas concerning the structure of the nucleons. It was found by
Burkardt and Jaffe \cite{Bur93} that the $u$ and $d$ quarks inside
a $\Lambda$ should be negatively polarized from SU(3) symmetry. It
was also pointed out by Soffer and one of us \cite{Ma99} that the
flavor and spin content of the $\Lambda$ can be used to test
different predictions concerning the spin structure of the nucleon
and the quark-antiquark asymmetry of the nucleon sea. Most
recently, we found \cite{MSY2} that the flavor and spin structure
of the $\Lambda$ near $x=1$ can provide clean tests between
perturbative QCD (pQCD) and the SU(6) quark-diquark model
predictions. We also found that the non-dominant up ($u$) and down
($d$) quarks should be positively polarized at large $x$, even
though their net spin contributions to the $\Lambda$ might be zero
or negative. Thus it is clear that the quark structure of
$\Lambda$ is a frontier which can enrich our understanding
concerning the flavor and spin structure of the nucleons and
provides a new domain to test various ideas concerning the hadron
structure that come from the available nucleon studies.

Unlike the nucleon case where the protons and neutrons (in the
nuclei) can be used as targets in various DIS processes, direct
measurement of the quark distributions of the $\Lambda$ is
difficulty, since the $\Lambda$ is a charge-neutral particle which
cannot be accelerated as incident beam and its short life time
makes it also difficult to be used as a target.
However, the quark distributions and the quark fragmentation
functions are interrelated quantities that can uncover the 
structure of the involved hadron \cite{Ja93,GLR}.  
For example, the quark distributions
inside a hadron are related by crossing symmetry to the
fragmentation functions of the same flavor quark to the same
hadron, by a simple reciprocity relation \cite{GLR}
\begin{equation}
q_{h}(x) \propto D_q^h(z),
\end{equation}
where $z=2 p \cdot q/Q^2$ is the momentum fraction of the produced
hadron from the quark jet in the fragmentation process, and
$x=Q^2/2 p \cdot q$ is the Bjorken scaling variable corresponding
to the momentum fraction of the quark from the hadron in the DIS
process. Although such a relation may be only valid
at a specific scale $Q^2$ for $x \to 1$ and $z \to 1$ under  
leading order approximation and there are corrections to this relation
from experimental observation and theoretical considerations
\cite{GLRc}, 
it can provide a reasonable connection
between different physical quantities and lead to different
predictions about the fragmentations based on our understanding of
the quark structure of a hadron \cite{Ma99,Bro97}. 
Among the
various possible hadrons that can be produced, $\Lambda$ hyperon
is most suitable for studying the polarized fragmentation due to
its self-analyzing property owing to the characteristic decay mode
$\Lambda \to p \pi^-$ with a large branching ratio of 64\%. Thus
we can use various $\Lambda$ fragmentation processes to
investigate the spin and flavor structure of the $\Lambda$ and to
test various ideas concerning the hadron structure.
From another point of view, studying the quark to $\Lambda$  
fragmentations
is also interesting in itself. We may consider our study
as a phenomenological method to parameterize the quark
to $\Lambda$ fragmentation functions,  and the validity 
and reasonableness of the
method can be checked by comparison with the experimental
data on various quark to $\Lambda$ fragmentation functions.   

There have been many proposals concerning the measurements of the
$\Lambda$ fragmentations functions in different processes, for
different physical goals [13-15,18-27], and in this paper we will
focus our attention on the longitudinally polarized case. One
promising method to obtain a complete set of polarized
fragmentation functions for different quark flavors is based on
the measurement of the helicity asymmetry for semi-inclusive
production of $\Lambda$ hyperons in $e^+e^-$ annihilation on the
$Z^0$ resonance \cite{Bur93}. Measurements of the light-flavor
quark fragmentations into $\Lambda$ have been also suggested from
polarized electron DIS process \cite{Jaf96} and neutrino DIS
process \cite{Kot98}, based on the $u$-quark dominance assumption.
It has been also suggested to determine the polarized
fragmentation functions by measuring the helicity transfer
asymmetry in the process $p \overrightarrow{p} \to
\overrightarrow{\Lambda} X$ \cite{Flo98}. From its dependence on
the rapidity of the $\Lambda$, it is possible to discriminate
between various parameterizations. There is also a recent
suggestion \cite{Ma99} to measure a complete set of quark to
$\Lambda$ unpolarized and polarized fragmentation functions for
different quark flavors by the systematic exploitation of
unpolarized and polarized $\Lambda$ and $\bar{\Lambda}$
productions in neutrino, antineutrino and polarized electron DIS
processes.

Recently there have been detailed measurements of the $\Lambda$
polarizations from the $Z$ decays in $e^+e^-$-annihilation
\cite{ALEPH96,DELPHI95,OPAL97}. The measured
$\Lambda$-polarization has been compared with several theoretical
calculations \cite{Kot98,Bor98,Flo98b} based on simple ansatz such
as $\Delta D_{q}^{\Lambda}(z) =C_{q}(z) D_{q}^{\Lambda}(z)$ with
constant coefficients $C_{q}$, or Monte Carlo event generators
without a clear physical motivation. It is the purpose of this
paper to calculate the $\Lambda$-polarization in
$e^+e^-$-annihilation at the $Z$-pole by using the physics results
presented in Ref.~\cite{MSY2}. It will be shown that the
quark-diquark model gives a very good description of the available
experimental data; pQCD can also give a good description of the
data by taking into account the suppression of quark helicities
compared to the SU(6) quark model values of quark spin
distributions. Thus the prediction of positive polarizations for
the $u$ and $d$ quarks inside the $\Lambda$ at $x \to 1$ is
supported by the available experimental data.

The paper is organized as follows. In Section II of the paper we
will present the formulas for the $\Lambda$-polarization in the
$e^+e^-$-annihilation near the $Z$-pole. In Section III we
calculate the $\Lambda$-polarization in the SU(6) quark-diquark
model and find that the model gives a very good description of the
available data. In Section IV we present the analysis for three
cases in the pQCD framework and find that we can also give a good
description of the data by taking into account the suppression of
quark helicities compared to the naive SU(6) quark model spin
distributions. Finally, we present discussions and conclusions in
Section V.

\section{$\Lambda$-Polarization in $e^+e^-$-Annihilation near the
$Z$-Pole}

One interesting feature of quark-antiquark ($q \bar q$) production
in $e^+e^-$-annihilation near the $Z$-pole is that the produced
quarks (antiquarks) are polarized due to the interference between
the vector and axial vector couplings in the standard model of
electroweak interactions, even though the initial $e^+$ and $e^-$
beams are unpolarized. Such quark (antiquark) polarization leads
to the polarization of the $\Lambda$ ($\bar{\Lambda}$) from the
decays of the quarks, therefore we can study the polarized quark
to $\Lambda$ fragmentations by the semi-inclusive production of
$\Lambda$ in $e^+e^-$-annihilation near the $Z$-pole
\cite{Bur93,Gus93,Bor98,Flo98b}.

The differential cross section for the $e^+e^- \to q \bar{q}$
process near the $Z$-pole is
\begin{equation}
\begin{array}{cllr}
\frac{d \sigma}{d \Omega} &=& N_c\frac{\alpha^2(Q^2)}{4 s} \left\{
(1+ \cos^2\theta) [e_q^2-2 \chi_1 v_e v_q e_q + \chi_2
(a_e^2+v_e^2) (a_q^2+v_q^2)] \right.
\\&+& \left.
2\cos\theta [-2 \chi_1 a_e a_q e_q+4 \chi_2 a_e a_q v_e v_q]
\right\},
\end{array}
\end{equation}
where
\begin{equation}
\chi_1=\frac{1}{16 \sin^2 \theta_W \cos^2 \theta_W}
\frac{s(s-M_Z^2)}{(s-M_Z^2)^2+M_Z^2\Gamma_Z^2},
\end{equation}
\begin{equation}
\chi_2=\frac{1}{256 \sin^4 \theta_W \cos^4 \theta_W}
\frac{s^2}{(s-M_Z^2)^2+M_Z^2\Gamma_Z^2},
\end{equation}
\begin{equation}
a_e=-1
\end{equation}
\begin{equation}
v_e=-1+4 \sin^2 \theta_W
\end{equation}
\begin{equation}
a_q=2 T_{3q},
\end{equation}
\begin{equation}
v_q=2 T_{3q}-4 e_q \sin^2 \theta_W,
\end{equation}
where $T_{3q}=1/2$ for $u$, $c$, while $T_{3q}=-1/2$ for $d$, $s$,
$b$ quarks, $N_c=3$ is the color number, $e_q$ is the charge of
the quark in units of the proton charge, $\theta$ is the angle
between the outgoing quark and the incoming electron, $\theta_W$
is the Weinberg angle, and $M_Z$ and $\Gamma_Z$ are the mass and
width of $Z^0$.

In the parton-quark model, the differential cross section for the
semi-inclusive hadron ($h$) production process $e^+e^- \to h+ X$
is obtained by summing over the above cross section, weighted with
the probability $D_q^h(z,Q^2)$ that a quark with momentum $P/z$
fragments into a hadron $h$ with momentum $P$,
\begin{equation}
\frac{d^2 \sigma^h}{d \Omega d z}=\sum\limits_q \frac{d \sigma}{d
\Omega} D_q^h(z,Q^2),
\end{equation}
where the $D_q^h(z,Q^2)$ are normalized so that
\begin{equation}
\sum\limits_h \int z D_q^h(z,Q^2) {\mathrm d} z=1.
\end{equation}
 The corresponding cross section for the production of
polarized hadron $h$ production can  be written as
\cite{Bur93}
\begin{equation}
\begin{array}{cllr}
\frac{d^2 \Delta \sigma}{d \Omega d z} &=& - N_c
\frac{\alpha^2(Q^2)}{2 s } \sum\limits_{q} \left\{ -e_q \chi_1
\{a_qv_e[\Delta D_q^h(z)-\Delta D_{\bar q}^h(z)](1+\cos ^2 \theta)
\right.
\\
&+& \left. 2 a_e v_q [\Delta D_q^h(z)+\Delta D_{\bar q}^h(z)] \cos
\theta \}\right.
\\
&+& \left. \chi_{2}\{(v_e^2+a_e^2)v_qa_q[\Delta D_q^h(z)-\Delta
D_{\bar q}^h(z)](1+\cos ^2 \theta) \right.
\\ &+& \left. 2 v_e a_e (v_q^2+a_q^2)[\Delta
D_q^h(z)+\Delta D_{\bar q}^h(z)] \cos \theta \} \right\}.
\end{array}
\end{equation}

The polarizations of the initial quarks from $e^+e^-$-annihilation
are given by
\begin{equation}
P_q=-\frac{A_q(1+\cos ^2 \theta )+B_q \cos \theta} {C_q(1+\cos ^2
\theta )+D_q \cos \theta},
\end{equation}
where
\begin{equation}
A_q=2 \chi_{2}(v_e^2+a_e^2)v_qa_q-2 e_q \chi_1 a_q v_e,
\end{equation}
\begin{equation}
B_q=4 \chi_2  v_e a_e (v_q^2+a_q^2)-4 e_q \chi_1 a_e v_q,
\end{equation}
\begin{equation}
C_q=e_q^2-2 \chi_1 v_e v_q e_q+ \chi_2 (a_e^2+v_e^2)
(a_q^2+v_q^2),
\end{equation}
\begin{equation}
D_q=8 \chi_2 a_e a_q v_e v_q-4 \chi_1 a_e a_q e_q.
\end{equation}
Averaging over $\theta$, one obtains $P_q=-0.67$ for $q=u$, $c$,
and $P_q=-0.94$ for $q=d$, $s$, and $b$ at the $Z$-pole. From the
cross section formulas for the unpolarized and polarized $h$
production, we can write the formula for the
$\Lambda$-polarization
\begin{equation}
P_{\Lambda}(\theta)=-\frac{\sum\limits_{q} \left\{A_q (1+\cos^2
\theta)[\Delta D_q^h(z)-\Delta D_{\bar q}^h(z)] + B_q \cos \theta
[\Delta D_q^h(z)+\Delta D_{\bar q}^h(z)]\right\}}{\sum\limits_{q}
\left\{C_q (1+\cos^2 \theta)[D_q^h(z)+D_{\bar q}^h(z)] + D_q \cos
\theta [ D_q^h(z)+ D_{\bar q}^h(z)]\right\}}.
\end{equation}
By averaging over $\theta$ we obtain
\begin{equation}
P_{\Lambda}=-\frac{\sum\limits_{q} A_q [\Delta D_q^h(z)-\Delta
D_{\bar q}^h(z)]}{\sum\limits_{q} C_q [D_q^h(z)+D_{\bar q}^h(z)]}.
\label{PL2}
\end{equation}

There have been measurements of the $\Lambda$-polarization near
the $Z$-pole \cite{ALEPH96,DELPHI95,OPAL97}. The ALEPH
collaboration \cite{ALEPH96} measured the $\Lambda$-polarization
by combining data of both $\Lambda$ and $\bar{\Lambda}$. Since the
$\bar{q}$ quark helicity is expected to be opposite that of the
$q$ quark, the identical polarization $P_{\Lambda}$ is assumed for
either $\Lambda$ and $\bar{\Lambda}$ in the treatment of the data.
Therefore we can consider their data as $P_{\Lambda}$ for
$\Lambda$ production by Eq.~(\ref{PL2}). From Eq.~(\ref{PL2}) it
can be found that we need both the quark and anti-quark
distributions to calculate the $\Lambda$-polarization
$P_{\Lambda}$ at all $x$. However, the main purpose of this paper
aims at checking the flavor and spin structure of the $\Lambda$ at
large $x$ predicted in Ref.~\cite{MSY2}, thus we neglect the
contribution from the sea quarks in our calculations of the
$\Lambda$-polarization in the following discussions.

\section{$\Lambda$-Polarization in the SU(6) Quark-Diquark Model }

Before we look into the details of the flavor and spin structure
for the valence quarks of the $\Lambda$, we briefly review the
analysis of the unpolarized and polarized quark distributions in
light-cone SU(6) quark-spectator-diquark model \cite{Ma96}, which can be
considered as a revised version of the original SU(6)
quark-diquark models \cite{DQM}. The light-cone formalism provides
a convenient framework for the relativistic description of hadrons
in terms of quark and gluon degrees of freedom \cite{LCF,LCr,BHL}.
Light-cone quantization has a number of unique features that make
it appealing, most notably, the ground state of the free theory is
also a ground state of the full theory, and the Fock expansion
constructed on this vacuum state provides a complete relativistic
many-particle basis for diagonalizing the full theory
\cite{Bro94}.

As we know, it is proper to describe deep inelastic scattering as
the sum of incoherent scatterings of the incident lepton on the
partons in the infinite momentum frame or in the light-cone
formalism. The unpolarized valence quark distributions $u_v(x)$
and $d_v(x)$ are given in this model by
\begin{eqnarray}
&&u_{v}(x)=\frac{1}{2}a_S(x)+\frac{1}{6}a_V(x);\nonumber\\
&&d_{v}(x)=\frac{1}{3}a_V(x), \label{eq:ud}
\end{eqnarray}
where $a_D(x)$ ($D=S$ for scalar spectator or $V$ for axial vector
spectator) is normalized such that $\int_0^1 {\mathrm d} x
a_D(x)=3$, and it denotes the amplitude for quark $q$ to be
scattered while the spectator is in the diquark state $D$. Exact
SU(6) symmetry provides the relation $a_S(x)=a_V(x)$, which
implies the valence flavor symmetry $u_{v}(x)=2 d_{v}(x)$. This
gives the prediction $F^n_2(x)/F^p_2(x)\geq 2/3$ for all $x$,
which is ruled out by the experimental observation
$F^n_2(x)/F^p_2(x) <  1/2$ for $x \to 1$. The SU(6) quark-diquark
model \cite{DQM} introduces a breaking to the exact SU(6) symmetry
by the mass difference between the scalar and vector diquarks and
predicts $d(x)/u(x) \to 0$ at $x \to 1$, leading to a ratio
$F_2^n(x)/F_2^p(x) \to 1/4$, which could fit the data and has been
accepted in most parameterizations of quark distributions for the
nucleon. It has been shown that the SU(6) quark-spectator-diquark
model can reproduce the $u$ and $d$ valence quark asymmetry that
accounts for the observed ratio $F_2^{n}(x)/F_2^{p}(x)$ at large
$x$ \cite{Ma96}. This supports the quark-spectator picture of deep
inelastic scattering in which the difference between the mass of
the scalar and vector spectators is essential in order to
reproduce the explicit SU(6) symmetry breaking while the bulk
SU(6) symmetry of the quark model still holds.

The quark helicity distributions for the $u$ and $d$ quarks can be
written as \cite{Ma96}
\begin{equation}
\begin{array}{llcr}
\Delta u_{v}(x)=u_{v}^{\uparrow}(x)-u_{v}^{\downarrow}(x)
=-\frac{1}{18}a_V(x)W_q^V(x) +\frac{1}{2}a_S(x)W_q^S(x);
\\
\Delta d_{v}(x)=d_{v}^{\uparrow}(x)-d_{v}^{\downarrow}(x)
=-\frac{1}{9}a_V(x)W_q^V(x),
\end{array}
\label{eq:sfdud}
\end{equation}
in which $W_q^S(x)$ and $W_q^V(x)$ are the Melosh-Wigner
correction factors \cite{Ma96,Ma91b,Ma98} for the scalar and axial
vector spectator-diquark cases. They are obtained by averaging
Eq.~(\ref{eqM1}) over ${\mathbf k}_{\perp}$ with $k^+=x {\cal M}$
and ${\cal M}^2=\frac{m^2_q+{\mathbf
k}^2_{\perp}}{x}+\frac{m^2_D+{\mathbf k}^2_{\perp}}{1-x}$, where
$m_D$ is the mass of the diquark spectator, and are unequal due to
unequal spectator masses, which leads to unequal ${\mathbf
k}_{\perp}$ distributions. The explicit expression for the
Melosh-Wigner rotation factor \cite{Ma91b} is
\begin{equation}
W_q(x,{\mathbf k}_{\perp}) =\frac{(k^+ +m)^2-{\mathbf
k}^2_{\perp}} {(k^+ +m)^2+{\mathbf k}^2_{\perp}} 
\label{eqM1},
\end{equation}
which ranges between $0 \to 1$ due to the quark intrinsic
transverse motions.
 From Eq.~(\ref{eq:ud}) one gets
\begin{eqnarray}
&&a_S(x)=2u_v(x)-d_v(x);\nonumber\\ &&a_V(x)=3d_v(x).
\label{eq:qVS}
\end{eqnarray}
Combining Eqs.~(\ref{eq:sfdud}) and (\ref{eq:qVS}) we have
\begin{eqnarray}
&&\Delta u_{v}(x)
    =[u_v(x)-\frac{1}{2}d_v(x)]W_q^S(x)-\frac{1}{6}d_v(x)W_q^V(x);
\nonumber    \\ &&\Delta d_{v}(x)=-\frac{1}{3}d_v(x)W_q^V(x).
\label{eq:dud}
\end{eqnarray}
Thus we arrive at simple relations \cite{Ma96} between the
polarized and unpolarized quark distributions for the valence $u$
and $d$ quarks. The relations (\ref{eq:dud}) can be considered as
the results of the conventional SU(6) quark model, and which
explicitly take into account the Melosh-Wigner rotation effect
\cite{Ma91b} and the flavor asymmetry introduced by the mass
difference between the scalar and vector spectators \cite{Ma96}.
The calculated polarization asymmetries $A_1^N=2 x
g_1^N(x)/F_2^N(x)$, including the Melosh-Wigner rotation, have
been found \cite{Ma96} to be in reasonable agreement with the
experimental data, at least for $x \geq 0.1$. A large asymmetry
between $W_q^S(x)$ and $W_q^V(x)$ leads to a better fit to the
data than that obtained from a small asymmetry.

One interesting feature predicted from the relation is that
$\Delta u(x)/u(x) \to 1$ and $\Delta d(x)/d(x) \to -1/3$ at $x \to
1$. The prediction $\Delta d(x)/d(x) \to -1/3$ is different from
the pQCD prediction $\Delta d(x)/d(x) \to 1$ at large $x$, and
with the available data it is still not possible to make a clear
distinction between the two predictions. Thus the $\Delta
d(x)/d(x)$ behavior at $x \to 1$ can provide a new test between
pQCD and the quark-diquark model predictions.

In the following we analyze the valence quark distributions of the
$\Lambda$ by extending the SU(6) quark-spectator-diquark model
\cite{Ma96} from the nucleon case to the $\Lambda$.
The $\Lambda$
wave function in the conventional SU(6) quark model is written as
\begin{equation}
|\Lambda^{\uparrow} \rangle =\frac{1}{2\sqrt{3}} [(u^{\uparrow} d
^{\downarrow} + d^{\downarrow} u ^{\uparrow}) -(u^{\downarrow} d
^{\uparrow} + d ^{\uparrow} u ^{\downarrow} )] s^{\uparrow} +
(\mathrm{cyclic ~~permutation}). \label{SU6}
\end{equation}
The SU(6) quark-diquark model wave function for the $\Lambda$ is
written as
\begin{equation}
\Psi^{\uparrow,\downarrow}_{\Lambda} =\sin \theta ~ \varphi_V |q V
\rangle ^{\uparrow,\downarrow} + \cos \theta ~ \varphi_S |q S
\rangle^{\uparrow,\downarrow},
\label{SU6L}
\end{equation}
with
\begin{equation}
\begin{array}{cllr}
|q V \rangle ^{\uparrow,\downarrow}&=&\pm
 \frac{1}{\sqrt{6}} [V_0(ds) u
^{\uparrow,\downarrow} - V_0(us) d ^{\uparrow,\downarrow} -
\sqrt{2} V_{\pm}(ds) u ^{\downarrow,\uparrow} + \sqrt{2}
V_{\pm}(us) d ^{\downarrow,\uparrow}];
\\
|q S \rangle ^{\uparrow,\downarrow}&=&\frac{1}{\sqrt{6}} [S(ds) u
^{\uparrow,\downarrow} + S(us) d ^{\uparrow,\downarrow} -2  S(ud)
s ^{\uparrow,\downarrow}],
\end{array}
\label{SU6LD}
\end{equation}
where $V_{s_z}(q_1 q_2)$ stands for a $q_1 q_2$ vector diquark
Fock state with third spin component $s_z$, $S(q_1 q_2)$ stands
for a $q_1q_2$ scalar diquark Fock state, and $\varphi_D$ stands
for the momentum space wave function of the quark-diquark with $D$
representing the vector (V) or scalar (S) diquarks. The angle
$\theta$ is a mixing angle that breaks the SU(6) symmetry at
$\theta \neq \pi/4$ and in this paper we choose the bulk SU(6)
symmetry case $\theta =\pi/4$.

 From Eq.~(\ref{SU6L}) we get the unpolarized quark distributions for the
three valence $u$, $d$, and $s$ quarks for the $\Lambda$,
\begin{equation}
\begin{array}{clcr}
&u_v(x)=d_v(x)=\frac{1}{4} a_{V(qs)}(x) +\frac{1}{12}
a_{S(qs)}(x);
\\
&s_v(x)=\frac{1}{3} a_{S(ud)}(x),
\end{array}
\end{equation}
where $a_{D(q_1 q_2)}(x) \propto  \int [\mathrm{d}^2
\vec{k}_\perp] |\varphi (x, \vec{k}_\perp)|^2$ ($D=S$ or $V$)
denotes the amplitude for the quark $q$ being scattered while the
spectator is in the diquark state $D$, and is normalized such that
$\int_0 ^1 a_{D(q_1 q_2)}(x) \mathrm{d} x =3$. We assume the $u$
and $d$ symmetry $D(qs)=D(us)=D(ds)$, from the $u$ and $d$
symmetry inside $\Lambda$.

We get from Eq.~(\ref{SU6L}) the spin distribution probabilities
in the quark-diquark model
\begin{equation}
\begin{array}{cllr}
&u^{\uparrow}_V=d^{\uparrow}_V=1/12;
&u^{\downarrow}_V=d^{\downarrow}_V=1/6;\\
&u^{\uparrow}_S=d^{\uparrow}_S=1/12;
&u^{\downarrow}_S=d^{\downarrow}_S=0;\\ &s^{\uparrow}_V=0;
&s^{\downarrow}_V=0; \\ &s^{\uparrow}_S=1/3;
&s^{\downarrow}_S=0;
\end{array}
\end{equation}
Similar to the nucleon case, the quark spin distributions for the
three valence quarks can be expressed as,
\begin{equation}
\begin{array}{clcr}
&\Delta u_v(x)=\Delta d_v(x)=-\frac{1}{12}
a_{V(qs)}(x)W_{V(qs)}(x) +\frac{1}{12} a_{S(qs)}(x)W_{S(qs)}(x);
\\
&\Delta s_v(x)=\frac{1}{3} a_{S(ud)}(x)W_{S(ud)}(x),
\end{array}
\end{equation}
where $W_D(x)$ is the correction factor due to the Melosh-Wigner
rotation and is expressed as
\begin{equation}
W_{D(q_1 q_2)}(x) =  \int [\mathrm{d}^2 \vec{k}_\perp] W_{D(q_1
q_2)}(x)(x,\vec{k}_\perp) |\varphi (x, \vec{k}_\perp)|^2/a_{D(q_1
q_2)}(x).
\end{equation}
One can turn off the Melosh-Wigner rotation effect by setting
$W_D(x)=1$, which should be only true at $x \to 1$. This case was
discussed in Ref.~\cite{MSY2}.

 In order to perform the calculation,
we employ the Brodsky-Huang-Lepage (BHL) prescription \cite{BHL}
of the light-cone momentum space wave function for the
quark-spectator
\begin{equation}
\varphi (x, \vec{k}_\perp) = A_D \exp \{-\frac{1}{8\alpha_D^2}
[\frac{m_q^2+\vec{k}_\perp ^2}{x} +
\frac{m_D^2+\vec{k}_\perp^2}{1-x}]\},
\end{equation}
with parameters (in units of MeV) $m_q=330$ for $q=u$ and $d$,
$m_s=480$, $\alpha_D=330$, $m_{S(ud)}=600$, $m_{S(qs)}=750$, and
$m_{V(qs)}=950$, following Ref.~\cite{Ma96}. The differences in
the diquark masses $m_{S(ud)}$, $m_{S(qs)}$, and $m_{V(qs)}$ cause
the symmetry breaking between $a_{D(q_1 q_2)}(x)$ in a way that
$a_{S(ud)}(x) > a_{S(qs)}(x) > a_{V(qs)}(x)$ at large $x$.

\begin{figure}[htb]
\begin{center}
\leavevmode {\epsfysize=5.5cm \epsffile{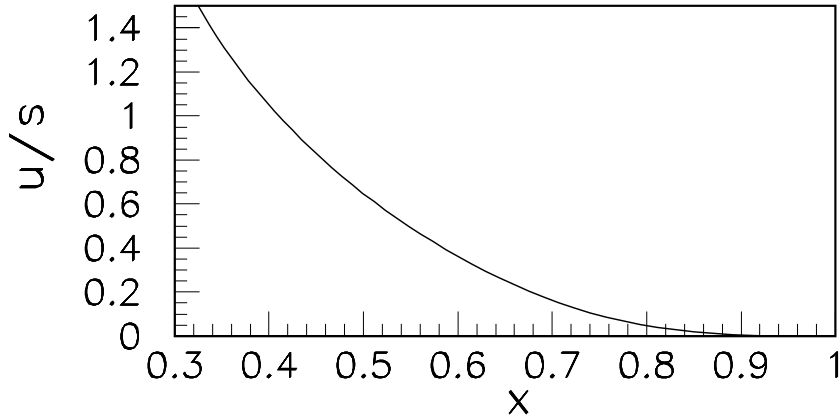}}
\end{center}
\caption[*]{\baselineskip 13pt The ratio $u(x)/s(x)$ of the
$\Lambda$ in the SU(6) quark-diquark model. }\label{msy3f1}
\end{figure}

\begin{figure}[htb]
\begin{center}
\leavevmode {\epsfysize=5.5cm \epsffile{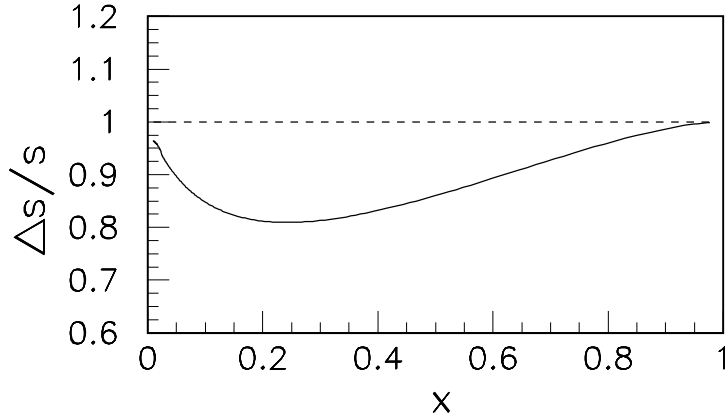}}
\end{center}
\caption[*]{\baselineskip 13pt The ratio $\Delta s(x)/s(x)$ for
the valence strange quark of the $\Lambda$ in the SU(6)
quark-diquark model. The solid and dotted curves are the
corresponding results with (solid) and without (dotted) the
Melosh-Wigner rotation.} \label{msy3f2}
\end{figure}

\begin{figure}[htb]
\begin{center}
\leavevmode {\epsfysize=5.5cm \epsffile{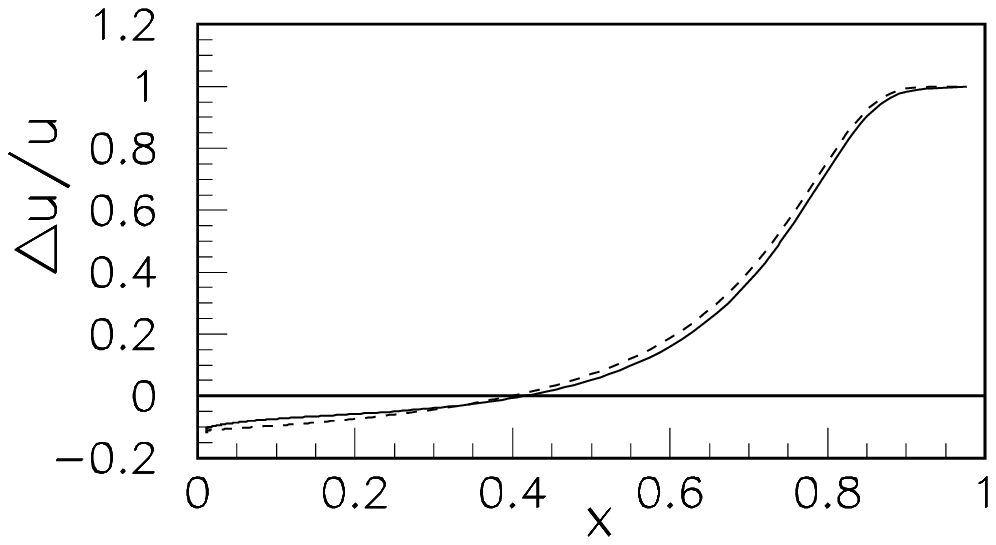}}
\end{center}
\caption[*]{\baselineskip 13pt The ratio $\Delta u(x)/u(x)$ for
the up and down valence quarks of the $\Lambda$ in the SU(6)
quark-diquark model. The solid and dotted curves are the
corresponding results with (solid) and without (dotted) the
Melosh-Wigner rotation.} \label{msy3f3}
\end{figure}

In Fig.~\ref{msy3f1} we present the ratio $u(x)/s(x)$ calculated
from the quark-diquark model. We also present in Fig.~\ref{msy3f2}
the ratio $\Delta s(x)/s(x)$ for the dominant valence $s$ quark
which provides the quantum numbers of strangeness and spin of the
$\Lambda$ , and the ratio $\Delta u(x)/u(x)$ for the non-dominant
valence $u$ and $d$ quarks. We find that the ratio $\Delta
s(x)/s(x)$ is not a constant equal to 1 as is the case without
Melosh-Wigner rotation. The ratio $\Delta u(x)/u(x)$, presented in
Fig.~\ref{msy3f3}, is also suppressed at $x \neq 1$. But the
end-point behaviors at $x \to 1$ is unchanged. Thus the
quark-diquark model predicts, in the limit $x \to 1$, that
$u(x)/s(x) \to 0$ for the unpolarized quark distributions, $\Delta
s(x)/s(x) \to 1$ for the dominant valence $s$ quark, and also
$\Delta u(x)/u(x) \to 1$ for the non-dominant valence $u$ and $d$
quarks.

\begin{figure}[htb]
\begin{center}
\leavevmode {\epsfysize=7.5cm \epsffile{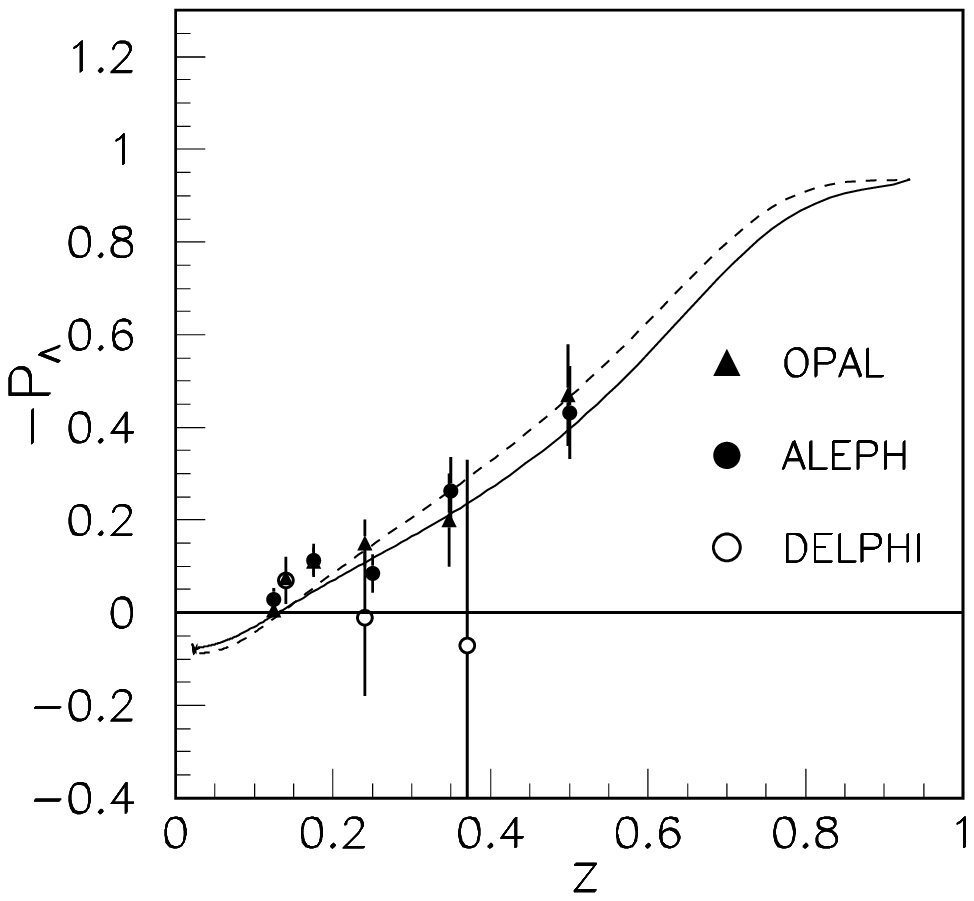}}
\end{center}
\caption[*]{\baselineskip 13pt The comparison of the experimental
data \cite{ALEPH96,DELPHI95,OPAL97} for the longitudinal
$\Lambda$-polarization $P_{\Lambda}$ in $e^+e^-$-annihilation
process at the $Z$-pole with the theoretical calculations in  the
SU(6) quark-diquark model. The solid and dotted curves are the
corresponding results with (solid) and without (dotted) the
Melosh-Wigner rotation.} \label{msy3f4}
\end{figure}

In Fig.~\ref{msy3f4} we present our calculated result for the
$\Lambda$-polarization $P_{\Lambda}(z)$ and we find that the
theoretical results from the quark-diquark model fit the data very
well within its present precision, at least in the large $z$
region. Thus the quark-diquark model provides a successful
description of the $\Lambda$-polarization $P_{\Lambda}(z)$, in
addition to its successful descriptions of the ratio
$F^n_2(x)/F^p_2(x)$ and the polarized structure functions for the
proton and neutron. It is necessary to point out that the
quark-diquark model with simple wave functions such as the BHL
prescription can provide good descriptions of the relations
between different quantities where the uncertainties in the model
can be canceled between each other. It is impractical to expect a
good description of the absolute magnitude and shape for a basic
physical quantity, such as the detailed feature of the cross
section, within such a model with simple wave functions.
In fact there have been 
calculations of the explicit shapes for the quark fragmentation
functions in a quark-diquark model \cite{Nza95} 
and for the quark distributions inside the $\Lambda$ 
in the MIT bag model \cite{Bor99}. We are very interested to notice
that the two works arrived at the same qualitative conclusion as ours
for a positive $u$ and $d$ polarization inside $\Lambda$ at large $x$ 
with small magnitude, though there are some difference in detailed
quantitative features.

\section{$\Lambda$-Polarization in pQCD Analysis}

We now look at the pQCD analysis of the quark distributions. In
the region $x \to 1$ pQCD can give rigorous predictions for the
behavior of distribution functions \cite{Bro95}. In particular, it
predicts ``helicity retention", which means that the helicity of a
valence quark will match that of the parent nucleon. Explicitly,
the quark distributions of a hadron $h$ have been shown to satisfy
the counting rule \cite{countingr},
\begin{equation}
q_h(x) \sim (1-x)^p, \label{pl}
\end{equation}
where
\begin{equation}
p=2 n-1 +2 \Delta S_z.
\end{equation}
Here $n$ is the minimal number of the spectator quarks, and
$\Delta S_z=|S_z^q-S_z^h|=0$ or $1$ for parallel or anti-parallel
quark and hadron helicities, respectively \cite{Bro95}. With such
power-law behaviors of quark distributions, the ratio $d(x)/u(x)$
of the nucleon was predicted \cite{Far75} to be 1/5 as $x \to 1$,
and this gives $F_2^n(x)/F_2^p(x)=3/7$, which is (comparatively)
close to the quark-diquark model prediction $1/4$. From the
different power-law behaviors for parallel and anti-parallel
quarks, one easily finds that $\Delta q/q =1$ as $x \to 1$ for any
quark with flavor $q$ unless the $q$ quark is completely
negatively polarized \cite{Bro95}. Such prediction are quite
different from the quark-diquark model prediction that $\Delta
d(x)/d(x)=-1/3$ as $x \to 1$ for the nucleon \cite{Ma96,Mel96}.
The most recent analysis \cite{Yang99} of experimental data for
several processes supports the pQCD prediction of the unpolarized
quark behaviors $d(x)/u(x)=1/5$ as $x \to 1$, but there is still
no definite test of the polarized quark behaviors $\Delta
d(x)/d(x)$ since the $d$ quark is the non-dominant quark for the
proton and does not play a dominant role at large $x$.

We extend the pQCD analysis from the proton case to the $\Lambda$.
From the SU(6) wave function of the $\Lambda$ we get the explicit
spin distributions for each valence quark,
\begin{equation}
\begin{array}{cllr}
&u^{\uparrow}=d^{\uparrow}=\frac{1}{2}; ~~~~
&u^{\downarrow}=d^{\downarrow}=\frac{1}{2};
\\
&s^{\uparrow}=1; &s^{\downarrow}=0.
\end{array}
\label{SU6q}
\end{equation}
In pQCD and at large $x$, the anti-parallel helicity distributions
can be neglected relative to the parallel ones, thus SU(6) is
broken to SU(3)$^{\uparrow}\times$SU(3)$^{\downarrow}$.
Nevertheless, the ratio $u^{\uparrow}/s^{\uparrow}$ is still $1/2$
\cite{Bro95}. Thus helicity retention implies immediately that
$u(x)/s(x) \to 1/2$ and $\Delta q(x)/q(x) \to 1$ (for $q=u$, $d$,
and $s$) for $x \to 1$, and therefore the flavor structure of the
$\Lambda$ near $x=1$ is a region in which accurate tests of pQCD
can be made.

 From the power-law behaviors of Eq.~(\ref{pl}), we write down a
simple model formula for the valence quark distributions,
\begin{equation}
\begin{array}{clcr}
q^{\uparrow}(x)   \sim x^{-\alpha}(1-x)^3; ~~~~ q^{\downarrow}(x)
\sim x^{-\alpha}(1-x)^5,
\end{array}
\label{pls}
\end{equation}
where $q^{\uparrow}(x)$ and $q^{\downarrow}(x)$ are the parallel
and anti-parallel quark helicity distributions and $\alpha$ is
controlled by Regge exchanges with $\alpha \approx 1/2$ for
nondiffractive valence quarks. This model is not meant to give a
detailed description of the quark distributions but to outline its
main features in the large $x$ region. We define $B_n=B(1/2,n+1)$
where $B(1/2,n+1)$ is the $\beta$-function defined by
$B(1-\alpha,n+1)=\int_0^1 x^{-\alpha}(1-x)^{n} {\mathrm d} x$ for
$\alpha=1/2$. Combining Eq.~(\ref{pls}) with Eq.~(\ref{SU6q}), we
get,
\begin{equation}
\begin{array}{cllr}
&u^{\uparrow}(x)=d^{\uparrow}(x)=\frac{1}{2B_3}x^{-\frac{1}{2}}(1-x)^3;
 ~~~~ &u^{\downarrow}(x)=d^{\downarrow}(x)=\frac{1}{2B_5}x^{-\frac{1}{2}}(1-x)^5;
\\
&s^{\uparrow}(x)=\frac{1}{B_3}x^{-\frac{1}{2}}(1-x)^3;
&s^{\downarrow}(x)=0,
\end{array}
\label{case1}
\end{equation}
which obviously satisfies that $u(x)/s(x)=1/2$ and $\Delta
q(x)/q(x)=1$ (for $q=u$, $d$ and $s$) as $x \to 1$, and it is easy
to find that $B_3=32/35$ and $B_5=512/693$.

However, the above simple model satisfies the SU(6) quark model
spin distributions, $\Delta s=1$ and $\Delta u=\Delta d=0$, and
the spin sum $\Sigma \Delta q=1$ which means that the helicity sum
of the quarks equals to the $\Lambda$ spin. From the nucleon case
we know that this is not true in the real situation and the quark
helicity sum is much more suppressed than the naive expectations
from the famous ``spin crisis" or ``spin puzzle"
\cite{SpinR,Bro88}. As emphasized in Ref.~\cite{Ma91b}, the
helicity distributions measured on the light-cone are related by
the Melosh-Wigner rotation to the ordinary spins of the quarks in
an equal-time rest-frame wave function description. Thus, due to
the non-collinearity of the quarks, one cannot expect that the
quark helicities will sum simply to the proton spin. From the
SU(3) symmetry argument of Burkardt-Jaffe \cite{Bur93}, we know
that the $s$ quark helicity $\Delta s=\int_0^1 \Delta s(x)
{\mathrm d} x$ is suppressed from the simple quark model value 1
to $\Delta s \approx 0.6$ and the $u$ and $d$ quarks are also
negatively polarized with quark helicities $\Delta u=\Delta d
\approx-0.2$. The reduction in the quark helicities might be from
sea quarks, but in this paper we simply assume that the
Burkardt-Jaffe values of quark helicities can be attributed to the
valence quarks, in order to amplify the effect due to the
reduction of the quark helicity distributions in the valence quark
region at large $x$ ($z$). For this purpose we adopt a more
general expression\footnote{The coefficients $A_q$, $B_q$, $C_q$
and $D_q$ ($q=u,d,s$) in this section are different from those in
section II.} for the quark distributions
\begin{equation}
\begin{array}{cllr}
&u^{\uparrow}(x)=d^{\uparrow}(x)=A_u x^{-\frac{1}{2}}(1-x)^3;
 ~~~~ &u^{\downarrow}(x)=d^{\downarrow}(x)=C_u x^{-\frac{1}{2}}(1-x)^5;
\\
&s^{\uparrow}(x)=A_s x^{-\frac{1}{2}}(1-x)^3;
&s^{\downarrow}(x)=C_s x^{-\frac{1}{2}}(1-x)^5,
\end{array}
\label{case2}
\end{equation}
with the following parameters
\begin{equation}
\begin{array}{cllr}
&A_u=0.4/B_3;
 ~~~~ &C_u=0.6/B_5;
 \\
&A_s=0.8/B_3;
 ~~~~ &C_s=0.2/B_5,
\end{array}
\label{case2p}
\end{equation}
which are fixed by the constraints,
\begin{equation}
s=\int_0^1 s(x) {\mathrm d} x=1; ~~~~u=d=\int_0^1 u(x){\mathrm d}
x =1, \label{qn}
\end{equation}
which is exact for the valence quarks due to the quark number
conservation, and
\begin{equation}
\Delta s=\int_0^1 \Delta s(x) {\mathrm d} x=0.6;  ~~~~\Delta
u=\Delta d=\int_0^1 \Delta u(x) {\mathrm d} x=-0.2,
\label{qhs}
\end{equation}
which should be strictly true only for total quark contributions
(valence+sea).  We consider Eq.~(\ref{case2}) as only a simplified
model case in order to check the effect of the quark helicity
suppression, but cannot be really true due to the absence of the
sea contributions. We find that the SU(6) large-$x$ relation
\begin{equation}
A_u=A_s/2 \label{su6r}
\end{equation}
is automatically satisfied for this case.

\begin{figure}[htb]
\begin{center}
\leavevmode {\epsfysize=7.5cm \epsffile{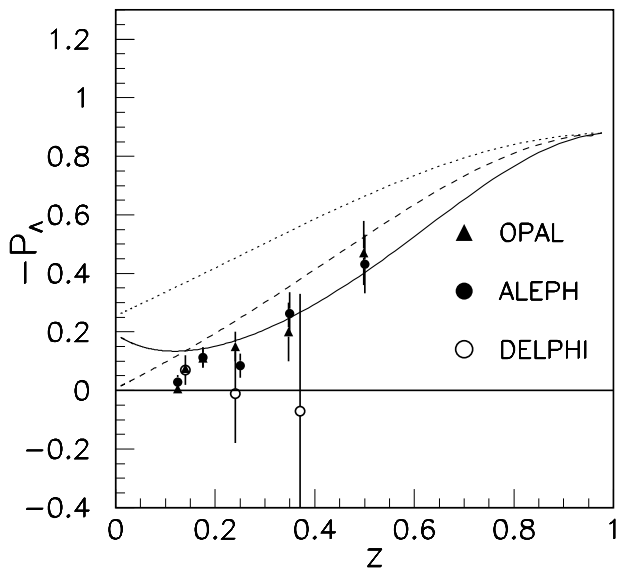}}
\end{center}
\caption[*]{\baselineskip 13pt The comparison of the experimental
data \cite{ALEPH96,DELPHI95,OPAL97} for the longitudinal
$\Lambda$-polarization $P_{\Lambda}$ in $e^+e^-$-annihilation
process at the $Z$-pole with the theoretical calculations in  the
the pQCD analysis with three different cases: (a) case 1: the
SU(6) quark-model spin distributions for the quark helicities,
Eq.~(\ref{case1}) (dotted curves); (b) case 2: the Burkardt-Jaffe
values for the quark helicities, Eq.~(\ref{case2}) (dashed curves)
; (c) case 3: the canonical form of quark distributions,
Eq.~(\ref{case3}) (solid curves). }\label{msy3f5}
\end{figure}

In Fig.~\ref{msy3f5} we present the calculated
$\Lambda$-polarization $P_{\Lambda}$ for the above two simple
cases of the pQCD analysis. For the case (case 1) of the naive
SU(6) quark model spin distributions for the quark helicities,
i.e., Eq.~(\ref{case1}), we find that the absolute magnitude is
larger than the experimental data and also than the previous
calculations \cite{Bor98,Flo98b}, which means that there should be
a source to reduce the quark helicities. The large magnitude of
$P_{\Lambda}$ is due to the large positive contributions from $u$
and $d$ quarks, i.e., positive $\Delta u(x)$, at large $x$ from
the pQCD prediction. It is interesting to find that in the case of
the Burkardt-Jaffe values of the valence quark helicities (case
2), i.e., Eq.~(\ref{case2}), one can describe the data well but
with a magnitude still slightly bigger, a result which is
different from previous calculations \cite{Bor98,Flo98b} in which
the reduction in the quark helicities causes much smaller
magnitudes of $P_{\Lambda}$ than the data. This means that the
reduction of quark helicities from the naive values of the SU(6)
quark spin distributions should provide a more physical picture
for the real world to describe the experimental data, contrary to
previous conclusions \cite{Bor98,Flo98b} that the naive SU(6)
quark model predictions fit the data better.

In fact, the above two simple pQCD cases still suffer from the
crudeness of the detailed shapes of the quark distributions with
only the leading term contributions. For a better reflection of
the complicated real situation we adopt the canonical form for the
quark distributions, following Ref.~\cite{Bro95},
\begin{equation}
\begin{array}{cllr}
&u^{\uparrow}(x)=&d^{\uparrow}(x)=A_u x^{-\frac{1}{2}}(1-x)^3+B_u
x^{-\frac{1}{2}}(1-x)^4;\\
&u^{\downarrow}(x)=&d^{\downarrow}(x)=C_u
x^{-\frac{1}{2}}(1-x)^5+D_u x^{-\frac{1}{2}}(1-x)^6;
\\
&s^{\uparrow}(x)=&A_s x^{-\frac{1}{2}}(1-x)^3+B_s
x^{-\frac{1}{2}}(1-x)^4;
\\
&s^{\downarrow}(x)=&C_s x^{-\frac{1}{2}}(1-x)^5+D_s
x^{-\frac{1}{2}}(1-x)^6.
\end{array}
\label{case3}
\end{equation}
 From the above constraints Eq.~(\ref{qn}),
we get
\begin{equation}
\begin{array}{clllc}
s&=&A_s B_3 +B_s B_4 +C_s B_5 +D_s B_6&=&1;
\\
u&=&A_u B_3 +B_u B_4 +C_u B_5 +D_u B_6&=&1;
\end{array}
\label{case3pa}
\end{equation}
where the $\beta$-functions $B_4=256/315$ and $B_6=2048/3003$.
From Eq.~(\ref{qhs}), we get
\begin{equation}
\begin{array}{clllc}
\Delta s&=&A_s B_3 +B_s B_4 -C_s B_5 -D_s B_6&=&0.7;
\\
\Delta u&=&A_u B_3 +B_u B_4 -C_u B_5 -D_u B_6&=&-0.1,
\end{array}
\label{case3pb}
\end{equation}
in which we have changed the valence quark helicities from the
Burkardt-Jaffe values $\Delta s=0.6$ and $\Delta u=-0.2$ to
$\Delta s=0.7$ and $\Delta u= -0.1$, to reflect the situation that
the sea quarks might contribute partially to the total $\Delta
s=0.6$ and $\Delta u=-0.2$. Combining with the SU(6) large-$x$
relation (\ref{su6r}) with Eqs.~(\ref{case3pa}) and
(\ref{case3pb}), we have only 5 constraints for the 8 parameters
and there are still large degrees of freedom to adjust the
parameters for a better fit of the $\Lambda$-polarization data.
For example, we choose $A_u=1/B_3$, $C_u=2/B_5$, and $C_s=2/B_5$
as inputs, and then have the following set of parameters:
\begin{equation}
\begin{array}{clllr}
&A_u=1/B_3; &B_u=-0.55/B_4; &C_u=2/B_5; &D_u=-1.45/B_6;
\\
&A_s=2/B_3; &B_s=-1.15/B_4; &C_s=2/B_5; &D_s=-1.85/B_6,
\end{array}
\label{case3b}
\end{equation}
which is denoted as case 3. This case is not meant to be totally
realistic but only to show that one can have a better description
of the available $P_{\Lambda}$ data with more reasonable picture
for the flavor and spin structure of the $\Lambda$. We also
present the results for this case in
Figs.~\ref{msy3f5}-\ref{msy3f8}. The ratios $\Delta s(x)/s(x)$ and
$\Delta u(x)/u(x)$ in this case have similar behaviors as those in
the quark-diquark model with the Melosh-Wigner rotation effect.
The calculated $\Lambda$-polarization $P_{\Lambda}$ can also give
a good description of the data at large $z$, as can be seen from
Fig.~\ref{msy3f5}, though the ratio of $u(x)/s(x)$ has the pQCD
behavior rather than the quark-diquark type, as seen by comparing
Fig.~\ref{msy3f6} with Fig.~\ref{msy3f1}. For the three cases of
pQCD analysis in our work, the ratios of $u(x)/s(x)$ for the
unpolarized quark distributions, $\Delta s(x)/s(x)$ for the
strange polarized quark distribution, and $\Delta u(x)/u(x)$ for
the $u$ and $d$ polarized quark distributions are presented in
Figs.~\ref{msy3f6}-\ref{msy3f8}. The quark momenta are also
calculated for the three cases and we find:
\begin{equation}
\begin{array}{clllr}
&\langle x_u \rangle=\langle x_d \rangle=0.094; &\langle x_s
\rangle=0.111; &\sum\limits_q \langle x_q \rangle=0.299, &{\mathrm
(case~1)}
\\
&\langle x_u \rangle=\langle x_d \rangle=0.091; &\langle x_s
\rangle=0.104; &\sum\limits_q \langle x_q \rangle=0.285, &{\mathrm
(case~2)}
\\
&\langle x_u \rangle=\langle x_d \rangle=0.118; &\langle x_s
\rangle=0.148; &\sum\limits_q \langle x_q \rangle=0.385. &{\mathrm
(case~3)}
\end{array}
\nonumber
\end{equation}

\begin{figure}[htb]
\begin{center}
\leavevmode {\epsfysize=5.5cm \epsffile{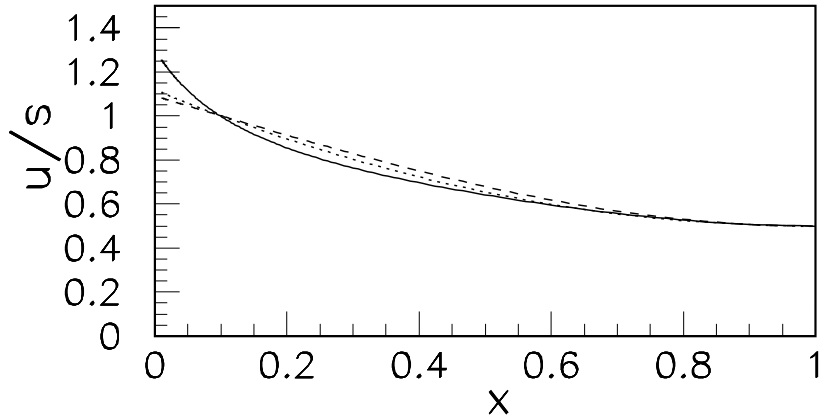}}
\end{center}
\caption[*]{\baselineskip 13pt The ratio $u(x)/s(x)$ of the
$\Lambda$ in the pQCD analysis with three cases: case 1 (dotted
curve); case 2 (dashed curve); and case 3 (solid curve).
}\label{msy3f6}
\end{figure}

\begin{figure}[htb]
\begin{center}
\leavevmode {\epsfysize=5.5cm \epsffile{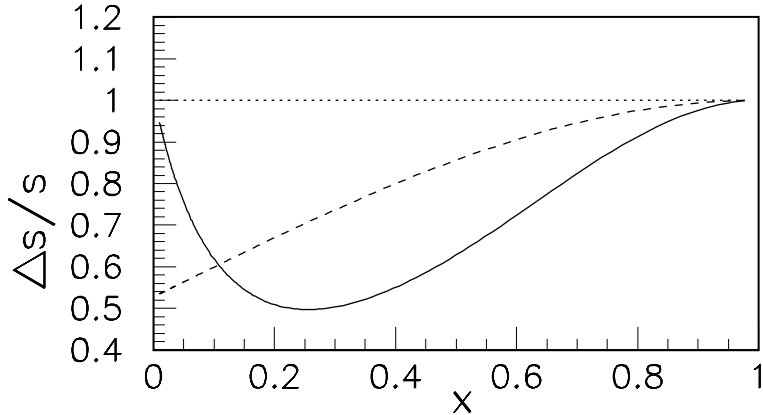}}
\end{center}
\caption[*]{\baselineskip 13pt The ratio $\Delta s(x)/s(x)$ for
the valence strange quark in the pQCD analysis with three cases:
case 1 (dotted curve); case 2 (dashed curve); and case 3 (solid
curve). }\label{msy3f7}
\end{figure}

\begin{figure}[htb]
\begin{center}
\leavevmode {\epsfysize=5.5cm \epsffile{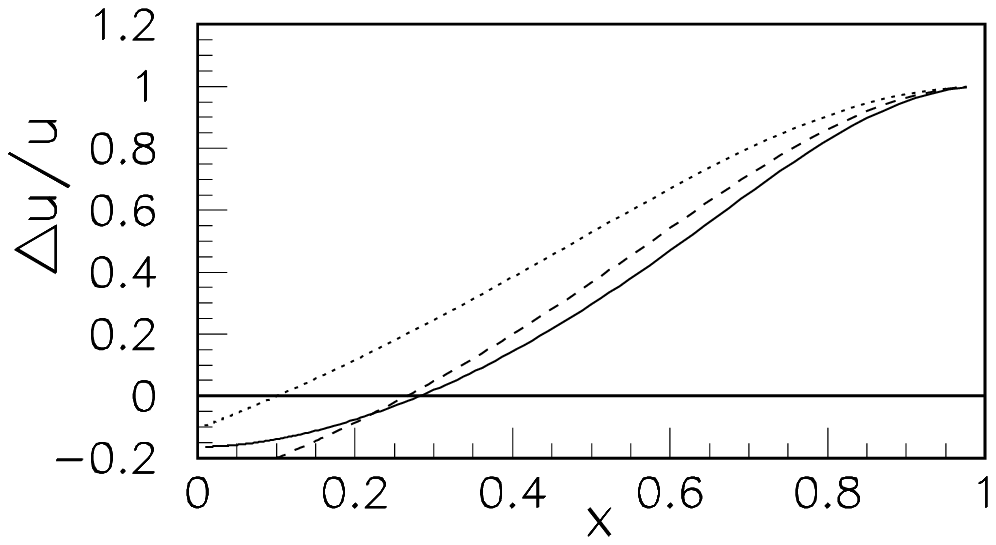}}
\end{center}
\caption[*]{\baselineskip 13pt The ratio $\Delta u(x)/u(x)$ for
the up and down valence quarks of the $\Lambda$ in the pQCD
analysis with three cases: case 1 (dotted curve); case 2 (dashed
curve); and case 3 (solid curve). }\label{msy3f8}
\end{figure}

It is interesting to notice that the ``most unlikely" scenario 3
in Ref.~\cite{Flo98b} is found to be better in reproducing the
data. From our work we know that their scenario 3 with all flavor
of quarks positively polarized is closer to our picture with
$\Delta q(x)/q(x)=1$ at $x \to 1$ for all quark flavors from the
pQCD analysis, thus it is not strange that this scenario can give
a better description of the data than the other two. But in the
pQCD analysis the net quark helicities for the valence $u$ and $d$
quarks should be zero or negative, which is different from their
scenario 3 in which the $u$ and $d$ quark helicities are positive.

There are other contributions that need to be considered
for a detailed description of the polarization. Those coming
from sea quarks and gluons have not been
considered in this work, and consequently the detailed features at small $x$
in Figs.~(\ref{msy3f4}) and (\ref{msy3f5}) should be
unreliable. The contributions of those $\Lambda$'s from the
decay of other hyperons have been discussed
and the corrections are found
to be small \cite{Gus93,Bor98}, therefore we can neglect them
as a first approximation.
In our
work the connection between the quark distributions and
the quark fragmentations should be only valid at low energy
scale of around a few GeV.
The evolution effects on the fragmentation
functions have been analyzed in Ref.~\cite{Flo98}, and from
re-producing the results in that work we notice that
the evolution has a
very small influence on the $\Lambda$ polarization.
Therefore the $\Lambda$ polarization from $e^+e^-$-annihilation
near the $Z$-pole
at the high energy scale does not alter the discussions concerning
the $\Lambda$ quark structure at the scale of our study.
Of course, all these contributions deserve further
study, along with the progress
of the experimental precision and a deeper understanding
concerning various quark to $\Lambda$ fragmentations.

\section{Discussions and Summary}

 From the above results in the paper, we found that the
quark-diquark model gives a very good description of the available
experimental data of the $\Lambda$-polarization in
$e^+e^-$-annihilation near the $Z$-pole. The pQCD analysis can
also describe the data well by taking into account the suppression
in the quark helicities compared to the naive SU(6) quark model
spin distributions. Unfortunately, it is still not possible to
make a clear distinction between the two different predictions of
the flavor and spin structure of the $\Lambda$ by only the
$\Lambda$-polarization in $e^+e^-$-annihilation near the $Z$-pole.
This can be easily understood since the quark polarizations,
$P_d=P_s=-0.94$ and $P_u=-0.67$, are close to each other, and the
same behaviors of $\Delta s(x)/s(x)$, $\Delta u(x)/u(x)$, and
$\Delta d(x)/d(x)$ near $x \to 1$ render it difficult to make a
clean separation of the contributions from different flavors. Thus
new information from other quantities related to the flavor and
spin structure of the $\Lambda$ are needed before we can have a
clean distinction between different predictions, and it seems that
$\Lambda$ ($\bar{\Lambda}$) production in the neutrino
(anti-neutrino) DIS processes \cite{Ma99} are more sensitive to
different flavors.

In summary, we studied the flavor and spin structure of the
$\Lambda$ at large $x$ in a pQCD analysis and in the quark-diquark
model, and then applied the results to discuss the
$\Lambda$-polarization of $\Lambda$ production in
$e^+e^-$-annihilation process near the $Z$-pole. We found that the
two theoretical frameworks give better description of the
available experimental data than previous calculations and also
provide more a reasonable picture, close to the real situation.
Thus the results in this paper can be considered as a
phenomenological support to our prediction \cite{MSY2} that the
$u$ and $d$ quarks should be positively polarized at large $x$,
even though their net helicities might be zero or negative. More
attention, both theoretically and experimentally, is needed to
study the flavor and spin structure of the $\Lambda$ for the
purpose of making a clear distinction between different
predictions.

{\bf Acknowledgments: }
This work is partially supported by
National Natural Science Foundation of China under Grant
No.~19605006, No.~19975052, and No.~19875024, and by Fondecyt (Chile)
postdoctoral fellowship 3990048, by Fondecyt (Chile) grant 1990806
and by a C\'atedra Presidencial (Chile).

\end{document}